\documentclass[12pt,a4paper]{article}

\usepackage[utf8]{inputenc}
\usepackage[T1]{fontenc}

\usepackage{amsmath}
\usepackage{amssymb}
\usepackage{amsfonts}
\usepackage{bm}

\usepackage{graphicx}
\usepackage{float}
\usepackage{subcaption}

\usepackage{booktabs}

\usepackage{multirow}

\usepackage[a4paper,margin=1in]{geometry}
\usepackage{setspace}
\usepackage{parskip}

\usepackage[colorlinks=true,
            linkcolor=blue,
            citecolor=blue,
            urlcolor=blue]{hyperref}

\usepackage[numbers,sort&compress]{natbib}

\title{\textbf{Construction and phenomenological analysis of dynamical Inverse seesaw neutrino mixing model}}

\author{%
  \small Manish Singh Bisht$^1$\href{https://orcid.org/0000-0001-8821-7439}{},
  Shuchi Bisht$^1$\href{https://orcid.org/0009-0003-2599-3192}{}, 
  Mayank Joshi$^1$\href{https://orcid.org/0009-0008-5001-7368}{}, 
  Navjot Hothi$^2$\thanks{hothi.navjot@gmail.com}~\href{https://orcid.org/0000-0001-7951-4212}{} \\[0.2cm]
  \small $^1$Department of Physics, D. S. B. Campus, Kumaun University,\\
  \small Nainital, Uttarakhand, India \\[0.1cm]
  \small $^2$Department of Physics, School of Engineering, University of Petroleum and Energy Studies,\\
  \small Dehradun, Uttarakhand, India
}

\date{}

\begin{document}

\maketitle

\begin{abstract}
In this manuscript, a dynamical inverse seesaw neutrino ISS(3+1+1+1) mixing model based on the gauge symmetry $U(1)_{B-L}$ extended by the flavor symmetry $A_{4}\times Z_{2}$ has been constructed. ISS(3+1+1+1) represents three generations of right-handed neutrinos and an extra three generation of singlets sterile neutrinos, wherein the right-handed neutrinos are triplets under $A_4$ symmetry. Furthermore, the five scalar flavon fields are considered in the model, which reproduce neutrino oscillation data within the NuFIT 6.0 experimental range with statistical $\text{|pull|}=0.01 \sigma$. The neutrino mass hierarchy is obtained by minimizing the dynamical $A_4\times Z_2$ scalar flavon $\phi_v$ potential. The effective neutrino mass parameter $m_{\beta\beta}$ for the neutrinoless double beta decay $0v\beta\beta$ has been charecterised in the manuscript. The effective neutrino mass parameter spans a finite region in the parameter space rather then being confined to a single point and it is constrained by the KamLAND-Zen excluded band. The model successfully explains active-sterile neutrino mixing and non-unitarity of the  PMNS matrix is consistent with the cosmological upper bound. The minimum value of the sum of the active neutrinos' mass turns out to be $m_{i}\approx 62\ \ (99.4)\ \text{meV}$ for Normal ordering (Inverted ordering), which lies within the range of both Planck 2018 loose cosmological limit $\sum m_{i}<120\ \text{meV}$ and tighter limit $\sum m_{i}<72\ \text{meV}$ of DESI-2024. Our model immensely favours normal ordering, in agreement with the tighter constraints reported by DESI-2024 .\\

\end{abstract}

\section{Introduction}

The Standard Model (SM) of particle physics gives a notably successfull comprehensive explanation of electromagnetic, weak, and strong interactions. Its success has attained the zenith after the discovery of Higgs boson\cite{denner2011standard,for2003search,bezrukov2008standard}. Nevertheless, the discovery of neutrino oscillations has indicated that neutrinos have small but non-zero masses and their flavors mix among themselves. It provides compelling experimental evidence for physics beyond the standard model (BSM)\cite{kajita2016nobel,farzan2018neutrino}. A minimal Standard Model does not have non-zero neutrino mass because its particle content exclude right handed neutrinos and also, it does not provide a mechanism for generating Majorana masses. The limitation of the minimal standard model motivates an extension of the standard model, which is able to generate  non-zero mass of the neutrinos and reproduce  experimentally observed lepton mixing parameters of the Pontecorvo-Maki-Nakagawa-Sakata (PNMS) matrix \cite{giganti2018neutrino, Pontecorvo1968, Maki1962}.\\
\noindent Several mechanisms have been proposed to generate smallness of the neutrino masses, which are classified as non-radiative and radiative methods. The non-radiative method includes type-I, type-II, type-III seesaw mechanisms\cite{type1,type2,type3} as well as inverse seesaw\cite{ISS}, linear seesaw mechanism\cite{LSS}. However, in contrast, radiative neutrino mass-generation mechanisms give small neutrino masses through quantum corrections to loop diagrams\cite{dror2020testing,bajc2007probing}. Among these, the ISS\cite{ISS} is the primary focus of this work, where the smallness of the active neutrino masses is determined by a small lepton-number-violating parameter $\mu$ rather than an ultrahigh mass scale or heavy neutral fermions mass that lie in the TeV scale\cite{dias2011inverse,karmakar20174}. The ISS framework encompasses a rich phenomenology linked with charged-lepton flavor violation, non-unitarity of the lepton mixing matrix, collider signature and the most important neutrinoless double beta decay ($0v\beta\beta$), contrary to conventional seesaw mechanisms of ultra high-scale counterparts\cite{ISS}.\\
\noindent The ISS explains the smallness of the neutrino masses but it does not explain the observed neutrinos mixing pattern\cite{XING2025116853}. Therefore, non-Abelian discrete symmetries ($S_3$, $S_4$, $A_4$, $A_5$, $\Delta(27)$) are widely used to constrain the Yukawa couplings\cite{nonab2,nonab}. The tetrahedral group $A_4$ \cite{zee} is the smallest group with a three-dimensional irreducible representation and thus naturally accommodates the three generations of leptons. Another well-motivated gauge extension arises from a global $U(1)_{B-L}$\cite{senoguz2005u} symmetry. With this extension, it is required to have right-handed neutrinos so that anomaly cancellation happens between gauge fields and lepton doublets. Spontaneous breaking of this symmetry generates masses for heavy neutral fermions and predicts the existence of an additional neutral gauge boson\cite{liu2022heavy,basso2009phenomenology}. \\
\noindent The notable experimental anomalies in the LSND\cite{athanassopoulos1997evidence}, MiniBooNE\cite{aguilar2021updated}, Gallium anti-neutrino\cite{krueger2023decoherence} and reactor anti-
neutrino experiments motivated the need for extension of the three
neutrino paradigm to a 3+1 , 3+2 , 3+1+1 active-sterile schemes\cite{sorel2004combined,peres20013+}. However, recent multi-collaboration data, notably Fermilab MicroBooNE's liquid-argon results, exclude the canonical light 3+1 sterile neutrino hypothesis at $95\%$ statistical confidence level\cite{soderberg2009microboone,karagiorgi2012microboone}. Therefore, the active-sterile schemes shift to alternatives, such as a heavy (keV-MeV) sterile state that decays in flight into a light neutrino plus a dark scalar or dark photon. This decay reproduces the low-energy electromagnetic signature without violating early-universe mass density limits.\\
 \noindent ISS based on the $A_4$ flavor symmetry have been studied widely\cite{karmakar20174,XING2025116853} in the literature. An extended $A_4$ have also been used\cite{borah2018a4} with additional cyclic symmetries can realise light Dirac neutrinos through both the type-I and the ISS. These models predict only NO, with the atmospheric mixing angle in the lower octant and $\delta_{CP}\in[-\pi/3,\pi/3]$. In contrast, a predictive radiative ISS model in the $3$-$3$-$1$ gauge framework~\cite{hernandez2018highly} with $A_4\times Z_4\times Z_6\times Z_{16}\times Z_{16}'$ symmetry excludes NO. It accommodates only IO, giving $m_{ee}=46.9~\text{meV}$ and $\delta_{CP}=-81.4^\circ$. Further realizations include ISS models based on non-holomorphic modular $A_4$ symmetry~\cite{zhang2025inverse}, as well as Dirac scoto-inverse seesaw models built on an accidental $U(1)_{B-L}$ symmetry that is spontaneously broken down to $\mathcal{Z}_3$. In a closely related minimal extended seesaw framework $A_4\times Z_3\times Z_2$ in a $3+1$ scheme, the lightest neutrino is massless~\cite{singh20253+}. In most of these works the mass ordering is fixed by the flavon alignment or by the assignment of irreducible representations, and when both orderings are allowed, they are fitted independently. In a more recent approach \cite{Capozzi2014}, a single real parameter has been introduced that smoothly interpolates between mixing patterns similar to NO and IO by entering as an additive term in the mass matrix. \\
\noindent A number of $A_4$ based ISS and gauged $U(1)_{B-L}$ ISS models exist in the literature \cite{singh20253+,zhang2025inverse}, these models generally employ different symmetry structures,  particle content or flavor assignments, like modular $A_4$\cite{singh20253+}, additional discrete factors or alternative gauge groups. The particular combination of the flavon-based $U(1)_{B-L}\times A_{4}\times Z_{2}$ symmetry with the field assignments, vacuum alignments, and the dynamical potential\cite{KarmakarSil2017} for ISS(3+1+1+1) structure are considered in the present work, which has not been investigated previously.\\
\noindent The present work differs in three respects: firstly, the connecting factor $\epsilon$ parametrises\cite{Hannestad:2016vhn} a continuous path in the vacuum alignment of $\Phi_\nu$. This consequently leads to the democratic and complex vacuum alignments, which simply does not require a correction after the mass matrix has been constructed. Secondly, the value of $\epsilon$ is not simply fitted minimisation of the $A_4\times Z_2$ scalar potential\cite{ishimori2010non,altarelli2010discrete}, rather $\epsilon=0$ or $\epsilon=1$ is chosen according to the sign of a single quartic coupling, so that the mass ordering is selected dynamically. Thirdly, all mass matrices $M_\ell$,
$M_D(\epsilon)$, $M_{RS}$ and $\mu$ are obtained by explicit $A_4$ Altarelli-Feruglio (A-F) tensor\cite{altarelli2010discrete} contraction rather than using a mere postulation. This fixes both their textures and reduces the model to twelve real parameters, which are considerably fewer than the constructions employing large products of cyclic groups. \\
\noindent The special features which distinguish the present work are that all mass matrices are properly derived and have not been simply assumed from the Lagrangian which has been constructed. The charged-lepton mass matrix $M_{l}$, the Dirac mass matrix $M_{D}(\epsilon)$, the heavy Dirac matrix $M_{RS}$ and the sterile matrix $\mu$ are generated by the explicit $A_{4}$ tensor contraction\cite{altarelli2010discrete} of the Lagrangian. This fixes the structure of the $\mu$ and $M_{RS}$ mass matrices, and each reduces to a two-parameter variable. It also fixes the charged lepton alignment. If the flavon generating $M_{RS}$ has a democratic alignment (i.e. same in all direction) the matrices $M_R$,$M_{RS}$ and $M_{D}$ mutually commute, and the $S_3$ permutation symmetry survives. The light neutrino mass matrix is then diagonalized by the discrete Fourier basis, which forces the exact tribimaximal mixing. Since the tribimaximal mixing is excluded by the reactor mixing data in 2012\cite{forero2012global}, the alignment $<\zeta>=(v_{\zeta},0,0)$ breaks the residual $S_3$ and restores the viable agreement with data. The right-handed Majorana mass term $v^{c}_{R}v_{R}$ is presented in the Lagrangian, but to restore the ISS condition, $|M_{R}|\ll |M_{RS}^{2}|/|\mu|$, this term mediates the active-sterile coupling, and decouples from the low energy scale for the light neutrino sector. The minimization of the renormalizable $A_{4}\times Z_{2}$ potential\cite{karagiorgi2012microboone} for the $\phi_v$ flavon field determines whether the global minimum lies at the democratic or the complex alignment. The effective Majorana mass depends heavily on two specific quantum phases\cite{Benato2015}, called "Majorana Phases". These two phases cancel each other in normal particle oscillation experiments and standard experiments and cannot place strict limits or boundaries on them. By fixing a total neutrino mass $(\sum m_i)$, the effective mass $m_{\beta\beta}$ magnifies more than ten times. Because of this huge variation, a single best-fit value for $m_{\beta\beta}$ is highly misleading. So instead of focusing on one single number, it is better to  look at the entire allowed range of possible values alongside the known limits from the cosmological bound range \cite{atre2009search}.\\
\noindent This paper is organized in five sections. Section \ref{2} presents the model description, which describes the particle content according to their group representation. Thereafter, the invariant Yukawa Lagrangian derives the charged-lepton and the neutrino mass matrices. Further, the ISS is implemented, and the effective light neutrino mass matrix is obtained. Section \ref{3} deals with the numerical analysis of neutrino masses, lepton mixing, CP violation, active-sterile mixing, non-unitarity, and neutrinoless double beta decay. Section \ref{4} provides the results of the numerical analysis along with the plots related to results of our model. Finally, conclusions are drawn in Section \ref{5} based upon model building by using the symmetry $U(1)_{B-L}\times A_{4}\times Z_{2}$.


\section{Theoretical Framework for Neutrino Mixing}\label{2}
We have analysed neutrino mass and mixing by using the dynamical ISS(3+1+1+1) supplemented by  $U(1)_{B-L} \times A_4 \times Z_2$ symmetry. The flavor structure is governed by the non-Abelian discrete symmetry $A_4 \times Z_2$, where the Standard Model lepton doublets $l_L$ are assigned as a triplet under $A_4$, and the right-handed charged leptons $(e_R,\mu_R,\tau_R)$ are assigned as different $A_4$ singlets such as 1, $1'$, $1''$ respectively. The right-handed neutrinos\cite{singh20253+,kishan20253+} are assigned as an $A_4$ triplet and the sterile neutrino are taken as $A_4$ singlets. The residual $Z_2$ symmetry plays a crucial role in stabilizing the lightest dark matter candidate. All Standard Model particles have $ Z_2$-even parity, and the non-Standard Model particles have $ Z_2$-odd parity. The gauge symmetry $U(1)_{B-L}$\cite{sanchez2014complex,singirala2018singlet} extends the standard model gauge group. This symmetry is broken by the vacuum expectation value of the scalar field $\eta$ with the $B-L$ charge.

For the invariant nature of the Lagrangian, the flavon fields $\phi$, $\phi_v$, $\zeta$, $\xi$ and $\eta$ have been introduced. The first three ($\phi$, $\phi_v$, $\zeta$) are the triplet under $A_4$ and the last two ($\zeta$ and $\xi$) are the $A_4$ singlet respectively. The role of these flavon fields $\phi$ and $\phi_v$ is to obtain the mass of the charged lepton and the mass of the active neutrinos, respectively, where the $\zeta$ is used for the $\nu_R - S$ coupling. The gauge group is $SU(3)_C\times SU(2)_L\times U(1)_Y\times U(1)_{B-L}$, extended by
$A_4\times Z_2$. Table-\ref{tab:first} provides the complete list of the full field content. \\
\begin{table}[H]
\centering
\caption{Particle content and their group representations.}\label{tab:first}
\begin{tabular}{| l | c | c | c | r|}
\hline
Field & $SU(2)_L\times U(1)_Y$ & $U(1)_{B-L}$ & $A_4$ & $Z_2$\\
\hline
$l_L$ & $(2,-1/2)$ & $-1$ & $\mathbf3$ & $+$\\
$e_R,\mu_R,\tau_R$ & $(1,-1)$ & $-1$ & $\mathbf1,\mathbf1'',\mathbf1'$ & $+$\\
$\nu_R$ & $(1,0)$ & $-1$ & $\mathbf3$ & $-$\\
$S$ & $(1,0)$ & $-1$ & $\mathbf1,\mathbf1'',\mathbf1'$ & $-$\\
$H$ & $(2,+1/2)$ & $0$ & $\mathbf1$ & $+$\\
$\Phi$ & $(1,0)$ & $0$ & $\mathbf3$ & $+$\\
$\Phi_\nu$ & $(1,0)$ & $0$ & $\mathbf3$ & $-$\\
$\zeta$ & $(1,0)$ & $0$ & $\mathbf3$ & $-$\\
$\xi$ & $(1,0)$ & $0$ & $\mathbf1$ & $+$\\
$\eta$ & $(1,0)$ & $+1$ & $\mathbf1$ & $-$\\
$\rho$ & $(1,0)$ & $-1$ & $\mathbf{1}$ & $-$ \\
\hline
\end{tabular}\end{table}

\noindent The invariant Lagrangian has been constructed based on the particle fields' contents and their group representation for the model:
\begin{equation}
\begin{aligned}
-\mathcal{L} &= \frac{y_e}{\Lambda^2}(\bar{\ell}_L H \Phi)_1 e_R \eta + \frac{y_\mu}{\Lambda^2}(\bar{\ell}_L H \Phi)_{1'} \mu_R \eta + \frac{y_\tau}{\Lambda^2}(\bar{\ell}_L H \Phi)_{1"} \tau_R \eta \\
& + \frac{y_2}{\Lambda^2}(\bar{\ell}_L H \Phi_\nu)_{3s} \nu_R \eta + \frac{y_1}{\Lambda^2}(\bar{\ell}_L H \Phi_\nu)_{3a} \nu_R \eta + \frac{y_3}{\Lambda^2}(\bar{\ell}_L H \nu_R)_1 \xi \eta \\
&+ \frac{y_R}{2\Lambda}(\xi^2 \overline{\nu_R^c} \nu_R + \Phi_\nu \xi \overline{\nu_R^c} \nu_R)\eta \\
&+ \frac{y_S}{\Lambda}((\nu_R S)_{3s} \zeta )_1 \xi \eta + \frac{y_S'}{\Lambda}((\nu_R S)_{3a} \zeta S)_1 \xi \eta \\
&+ \frac{\mu_S}{2}(\overline{S^c} S)_1 \eta + \frac{\mu_S'}{2}(\overline{S^c} S)_{1'} \eta +V(\eta,\rho)+\text{H.c.}
\end{aligned}
\end{equation}
where $k= v_{H} v_{\eta}v_{\zeta}/\lambda^{2}= 1.628\times 10^{5}\ \text{GeV}$ and the scalar $U(1)_{B-L}$ potential  $V(\eta,\chi)$ \cite{KarmakarSil2017} is defined as: 
\begin{equation}
    V(\eta,\rho)= \mu_{\eta}^{2}\eta^{\dagger}\eta + \mu_{\rho}^{2}\rho^{\dagger}\rho + \lambda_{\eta}(\eta^{\dagger}\eta)^{2} +\lambda_{\rho}(\rho^{\dagger}\rho)^{2}+ \lambda_{\eta\rho}(\eta^{\dagger}\eta)(\rho^{\dagger}\rho)
\end{equation}
 Yukawa couplings $y_1$, $y_{2}$ and $y_3$ \cite{mohapatra1980local,singh20253+} are complex. The subscripts $3s$ and $3a$ denote the symmetric and antisymmetric tensor products under the $A_4$ triplet contraction rule and $\Lambda$ represents the cutoff scale. Furthermore, by choosing the specific vacuum alignment $<\phi>= (v_{\phi},v_{\phi},v_{\phi})$ for the charged lepton sector and applying the Altarelli–Feruglio single contraction rule\cite{altarelli2010discrete}, the first three terms of the Lagrangian give $(l\phi)_{1}=v_{\phi}l_{1}$, $(l\phi)_{1'}=v_{\phi}l_{2}$ and $(l\phi)_{1"}=v_{\phi}l_{3}$, thereby producing a diagonal charged lepton mass matrix;
\noindent The flavon field $\phi_v$ in the Dirac neutrino sector takes the specific vacuum alignment, where it connects the two different linear term by an interpolation factor $\epsilon$\cite{Hannestad:2016vhn}, such as:
\begin{equation}
< \phi_{v}(\epsilon)>=(1-\epsilon)(v,v,v)+\epsilon\,(v,0,v\omega)
=\big(v,\;v(1-\epsilon),\;v[1+\epsilon(\omega-1)]\big)
\label{equ4}
\end{equation}
where $\omega= e^{i\frac{2\pi}{3}}$ and $\epsilon = 0\ \text{or}\ 1$. This reproduces $(v,v,v)$ at $\epsilon=0$, leading to the vacuum alignment for NO, and $(v,0,v\omega)$ at $\epsilon=1$, leading the vacuum alignment for IO. It must be noted that $<\xi>=(v_{\xi},0,0)$ is a singlet and is different from $\phi_v$ due to $A_4$ contraction \cite{altarelli2010discrete}. This specific vacuum alignment produces the Dirac mass matrix:
\begin{equation}
M_D(\epsilon)=k\begin{pmatrix}
2v_1y_2+y_3 & v_2A & -v_3B\\
v_3A & -v_1B & 2v_2y_2+y_3\\
-v_2B & 2v_3y_2+y_3 & v_1A\end{pmatrix}
\label{eq:MD}
\end{equation}
where $A= y_1 - y_2$ and $B= y_1 +y_2$. The property of the Dirac mass matrix is related to the value of $\epsilon$, therefore, for $\epsilon= 0$, the Dirac mass matrix produces exact circulant structure. The diagonal component of the Dirac mass matrix is independent from $\epsilon$ because $y_3$ couples with a singlet $\xi$ not $\phi_v$. For $\epsilon=1$, the component of the Dirac mass matrix $(M_{D})_{12}=(M_{D})_{31}=0$, which breaks the circulant form of the matrix.

\noindent Now, if we go for the solution of the heavy Dirac ($v_{r}-S$) neutrino mass matrix, then the democratic vacuum alignment of $<\zeta>= (v_{\zeta},v_{\zeta},v_{\zeta})$ gives the exact tribimaximal structure. The mixing angle is ruled out by the reactor mixing data because, in that case, residual symmetry $S_3$ appears\cite{HARRISON200376} after the breaking of the higher symmetry and the mass matrix $M_D$, $M_{RS}$ and $\mu $ commutes mutually. Therefore, the heavy neutrino($v_{r}-S$) mass matrix $M_{RS}$ can be defined as $M_{RS}=m_{RS}\big(T_S+\kappa\,T_A\big)$, where $T_S$ and $T_A$ are the symmetric and the antisymmetric matrices respectively. This generates $M_{RS}$ to be anti-symmetric in nature, and for the specific vacuum alignment for $<\zeta>=(v_{\zeta},0,0)$, $T_s$ and $T_A$ are defined as:
\begin{equation}
T_S=
\begin{pmatrix}
2 & 0 & 0 \\
0 & 0 & -1 \\
0 & -1 & 0
\end{pmatrix} \quad
T_A=\begin{pmatrix}
0 & 0 & 0 \\
0 & 0 & 1 \\
0 & -1 & 0
\end{pmatrix}
\end{equation}
Where $M_{RS}^{T}=-M_{RS}$ with $\kappa = y_{S}'/y_{S}$. Futhermore, $\text{det}\ M_{RS}=2 m_{RS}^{3}(\kappa^{2}-1)$, therefore the term $k=\pm 1$ must be excluded from the solution. The physical interpretation of $k=\pm 1$ may be associated with the possibility that different types of sterile neutrinos exist in nature, characterized by distinct interaction properties with other particles or among themselves.  

\noindent In the obtained Lagrangian, there are three sterile neutrinos $A_4$ singlets mass terms, which are obtained from A-F contraction rule\cite{altarelli2010discrete}, where $\mu_{ij}$ is in the bilinear form, which is obtained by extracting them using the equation $\mu_{ij}=\tfrac12\,\partial^2\mathcal{L}/\partial S_i\partial S_j$, and it ultimately gives:
\begin{equation}
\mu_{ij}=\begin{pmatrix}\mu_S&\mu_S'&0\\\mu_S'&0&\mu_S\\0&\mu_S&\mu_S'\end{pmatrix} 
\qquad \text{and} \qquad
\det\mu_{ij}=-(\mu_S+\mu_S')\big(\mu_S^2-\mu_S\mu_S'+\mu_S'^2\big)
\label{eq:mu}
\end{equation}
The matrix obtained above is a complex symmetric and invertible, with two complex (four real) parameters.\\
\noindent Finally, ISS will be used to determine the active neutrino mass\cite{dias2011inverse,karmakar20174}. After spontaneous symmetry breaking, we get $9\times 9$ neutrino mass matrix in the basis of $\psi^{T}=(v_{L},v_{R}^{c},S)$ given as:
\begin{equation}
\mathcal{M} = \begin{pmatrix}
0 & M_D & 0 \\
M_D^T & 0 & M \\
0 & M^T & \mu
\end{pmatrix}
\end{equation}
Here, the right handed Majorana neutrino $M_R$ matrix term has been removed because $M_R \gg M_{RS}^{2}/\mu^{2}\approx 10^{15}\ \text{GeV}$, which is close to the GUT energy scale. At the low energy scale, it decouples with the light neutrinos. Therefore, applying the condition of ISS $\mu\ll M_{D}\ll M_{RS}$, we can generate the effective light neutrino mass matrix. The $9\times 9$ neutrino mass matrix  diagonalizes as:
\begin{equation}
m_\nu = M_D M_{RS}^{-1} \mu(M_{RS}^{-1})^{T} M_D^T
\end{equation}
where the heavy neutrino mass matrix is $M_{RS}$. Thus, this exercise has been performed on the basis of the ISS model. In the ISS mechanism, the masses of the active neutrinos are obtained without any need for the heavy Majorana right-handed neutrino. However, the heavy Majorana right-handed interaction term is still present in the Lagrangian. This term plays a crucial role in the ISS framework where it mediates the coupling between the active and the sterile neutrinos. 


\subsection{Vacuum potential for dynamical ordering }
The flavon field $\phi_v$ is responsible for the vacuum structure of the model through its scalar potential. Two candidates for vacuum alignments that are most promising to lead to phenomenologically preferred neutrino mass ordering are considered: democratic vacuum alignment $(v,v,v)$ and complex vacuum alignment $(v,0,v\omega)$ (where $\omega=e^{2\pi i/3}$). We obtain these vacuum alignments by minimizing most general renormalizable scalar potential which is allowed by $A_4\times Z_2$ symmetry. As $\phi_v$ is $A_4$ triplet and odd under $Z_2$, the corresponding potential will be given by all renormalizable terms invariant under $A_4\times Z_2$ symmetry and can be written as\cite{karmakar20174}:
\begin{equation}
V=-\mu^2(\phi_v\phi_v)_{\mathbf1}
+\lambda_1\big[(\phi_v\phi_v)_{\mathbf1}\big]^2
+\lambda_2(\phi_v \phi_v)_{\mathbf1'}(\phi_v \phi_v)_{\mathbf1''}
+\lambda_3\,I_{SS},
\label{eq:V}
\end{equation}
Here
$I_{SS}\equiv\big[(\phi_v\phi_v)_{\mathbf3_S}(\phi_v\phi_v)_{\mathbf3_S}\big]_{\mathbf1}$ is the allowed fourth symmetric triplet contraction product.
The invariants take the values $\{3,9,0\}$ at $(v,v,v)$ and
$\{1,2,-4\}$ at $(v,0,v\omega)$. Therefore, it is obvious that the ISS affect the inverse ordering energy but leaves the NO energy term. For the extremum $\partial V/\partial \phi =0$,  the deepest depth for the both cases have to be measured. Therefore
\begin{equation}
\mu^2_{\rm NO}=6v^2(\lambda_1+\lambda_2),\qquad
\mu^2_{\rm IO}=2v^2(\lambda_1+2\lambda_2-4\lambda_3),
\end{equation}
and comparing the depths yields:
\begin{equation}
\begin{split}
\text{NO global minimum}\iff4\lambda_1+5\lambda_2-4\lambda_3>0,
\\
\text{IO global minimum}\iff4\lambda_1-\lambda_2+20\lambda_3>0.
\label{eq:cond}
\end{split}
\end{equation}
It is important to note that $I_{ss}$ term vanishes at $(v,v,v)$ for the NO alignment but equals $-4$ at the IO alignment. Therefore $\lambda_3$ acts as a pure ordering switch with critical value;
\begin{equation}
\lambda_3^{\rm crit}=\frac{4\lambda_1+5\lambda_2}{4}.
\end{equation}
The interpolating factor is therefore not a fit parameter, and it is fixed to $\epsilon=0$ or $\epsilon=1$, which depends upon the sign of
$\lambda_3-\lambda_3^{\rm crit}$.


\section{Numerical Analysis}\label{3}
In this section, numerical analysis is performed to scan the allowed parameter space for the model by taking into account the complex Yukawa couplings, the vacuum alignment, and the chosen cutoff scale to compute the active neutrino mass matrix via the ISS formula. The main focus oSf this analysis is to check whether the resulting neutrino masses and mixing angles, as well as the CP-violating phases \cite {mohapatra1980local}, fall within the experimentally favored ranges from global fits of oscillation data. Simultaneously, the preservation of the main theory of the ISS mechanism is sought, i.e., the smallness of the active neutrino masses is controlled by a small lepton number violating parameter $\mu$ rather than by extremely high mass scale heavy neutral fermions. By performing this numerical analysis, the aim is to establish that the $U(1)_{B-L}\times A_4\times Z_2$ framework can offer a viable and testable explanation of the observed neutrino oscillation phenomena. The complex Yukawa coupling for the Dirac neutrino mass term containing a phase term can be written as:
\begin{equation}
y_i=|y_i|e^{i\phi_i},
\qquad i=1,2,3.
\label{eq:complex_yukawa}
\end{equation}
The complex phases pertain to the non-trivial leptonic CP-violation term. The phases of the Yukawa couplings contribute to the physical Dirac CP phase and the Majorana phase after diagonalization of the light neutrino mass matrix. The Yukawa couplings are restricted to the weakly perturbative region, $10^{-3}\lesssim |y_i| \lesssim 10^{-2}$.\\
\noindent In this neutrino mass model, there have been 12 real free parameters: $m_{RS}$, $\kappa$, the
complex $y_{1,2,3}$ (six real) and the complex $\mu_S,\mu_S'$ (four
real). We have diagonalized the complex symmetric $m_\nu$ by Takagi
decomposition\cite{Takagi1990} by forming $H=m_\nu^\dagger m_\nu$ and then diagonalizing it.
The diagonalization of $H=U_0\,\mathrm{diag}(\lambda_i)\,U_0^\dagger$, taking
$\sigma_i=\sqrt{\max(\lambda_i,0)}$, and applying the phase correction
$U_{:,i}\to U_{:,i}e^{-\frac{i}{2}\arg(D_{ii})}$ with $D=U_0^Tm_\nu U_0$
so that $U^Tm_\nu U=\mathrm{diag}(m_1,m_2,m_3)$ is real and
non-negative. Takagi decomposition returns ascending singular values while the lightest
state is $m_1$ for NO and $m_3$ for inverse ordering (IO). For these realization the labels and the columns of $U_{PMNS}$ have been permuted
\begin{equation}
\text{NO}:(m_1,m_2,m_3)=(\sigma_1,\sigma_2,\sigma_3),\qquad
\text{IO}:(m_1,m_2,m_3)=(\sigma_3,\sigma_1,\sigma_2),
\label{eq:relabel}
\end{equation}
which makes $\Delta m^2_{31}<0$ for IO. From the PMNS mixing matrix $U$\cite{giganti2018neutrino,Pontecorvo1968,Maki1962} we have obtained:
\begin{equation}
 \sin^2\theta_{13}=|U_{e3}|^2 ;\quad    \sin^2\theta_{12}=\frac{|U_{e2}|^2}{1-\sin^2\theta_{13}}; \quad  \sin^2\theta_{23}=\frac{|U_{\mu3}|^2}{1-\sin^2\theta_{13}}  
 \end{equation}
the mixing angles. Also from the PMNS mixing matrix the Dirac CP-violation phase $\delta_{CP}$ via the standard $\delta_{CP}=\mathrm{atan2(sin\delta_{CP},cos\delta_{CP}})$ mod $360^{\circ}$  and the Jarlskog invariant $J_{CP}=\mathrm{Im}[U_{e1}U_{\mu2}U_{e2}^*U_{\mu1}^*]$ have been obtained. Furthermore, the effective Majorana mass $m_{\beta\beta}=|U_{e1}^2m_1+U_{e2}^2m_2+U_{e3}^2m_3|$, Majorana phases $\alpha_{21}=arg(U_{e2}^{2}m_2)-arg(U_{e1}^{2}m_1)$ and $\alpha_{31}=arg(U_{e3}^{2}m_3)-arg(U_{e1}^{2}m_1)$, active-sterile mixing term $\Theta=M_DM_{RS}^{-1}$ and non-unitarity factor $\eta=\tfrac12\Theta\Theta^\dagger$ have been obtained.

For minimization, we use $\chi^2=\sum_O\big[(O-O_{\rm target})/\sigma_O\big]^2+\sum_kP_k$, where $O$ runs over the six NuFIT~6.0 observables, treating $\delta_{CP}$ circularly so that the phase difference is $360^{\circ}$. The penalties term $P_k$ enforces $|y_i|\in[10^{-3},10^{-2}]$, $\sigma(\mu)\in[10^{-6},10^{-4}]$~GeV, $\sigma(M_D)\in[10^{2},10^{3}]$~GeV, $\sigma(M_{RS})\in[10^{5},10^{6}]$~GeV, the hierarchy $\max\sigma(\mu)<\min\sigma(M_D)<\min\sigma(M_{RS})$ and $\sum m_i$ is the cosmological upper limit, which fixes the absolute mass scale. This means that oscillation data determine only two mass-squared splittings; it is the cosmological limit which fixes the absolute mass scale and the Majorana mass $m_{\beta\beta}$. The value of the cosmological limit has been tightened in recent years; therefore, its value is treated as an external input rather than taken as a single value, which repeats the best-fit value across the range spanned by Planck 2018\cite{planck2020planck} and DESI 2024\cite{adame2025desi}.\\
Minimisation uses differential evolution (population 20, up to 500 generations) followed by L-BFGS-B polishing. The procedure
reaches $\chi^{2}=2.7\times10^{-8}$, $4.0\times10^{-5}$ and
$1.0\times10^{-9}$ for solutions~A, B and~C respectively. Therefore, all six observables are reproduced well inside their $0.01\sigma$ ranges. All group-theoretic derivations are performed symbolically and verified independently. 
\noindent  The entire numerical analysis has been performed in Python $3.12$ using NumPy $2.4$ for linear algebra (Takagi decomposition via Hermitian eigen decomposition, singular-value decompositions) and SciPy $1.17$ for optimisation. Global minimisation of the $\chi^{2}$ used differential evolution\cite{price2005differential} with population size 20 and up to 500 generations. This is followed by L-BFGS-B polishing, repeated over 8-12 independent random seeds. All group-theoretic derivations like the $A_{4}$ tensor contractions yielding $M_D$, $M_{RS}$, $\mu$ and $M_{l}$, and the scalar-potential minimisation conditions have been carried out symbolically with SymPy 1.14 and verified independently. Figures have been produced with Matplotlib 3.10.\\


\section{Results}\label{4}
\subsection{Best fit solution}
On the basis of the current cosmological bounds, there are three distinct solutions. The solutions A and B are obtained with the loose Planck 2018\cite{planck2020planck} $\sum m_{i}< 120\ \text{meV}$ for normal and IO, respectively, and C is obtained with the tighter DESI-2024\cite{Allali2024DESI} value $\sum m_{i}<72\ \text{meV}$.
\begin{table}[H]
\centering
\caption{Fitting parameters for neutrino observables. Column A corresponds to the loose Planck\cite{planck2020planck} bound $\sum m_{i}< 120\ \text{meV}$ for both normal and IO, whereas Column B reveals the parameter constraints for the tighter DESI \cite{Allali2024DESI} $\sum m_{i}<72\ \text{meV}$, for which IO is excluded.}\label{tab:params}
\begin{tabular}{|l|c|c|r |}
\hline
Parameters & A: NO ($120$) & A: IO ($120$) & B: NO ($72$)\\
\hline
$\chi^2_{min}$ & $2.73\times 10^{-8}$ & $3.97\times 10^{-5}$ & $1.03\times 10^{-9}$\\
$\epsilon$ (from potential) & $0$ & $1$ & $0$\\
$m_{RS}$ [GeV] & $1.333\times 10^5$ & $1.285\times 10^5$ & $3.186\times 10^5$\\
$\kappa=y_S'/y_S$ & $-2.009$ & $-2.963$ & $-1.732$\\
$|y_1|$ & $1.000\times 10^{-3}$ & $1.029\times 10^{-3}$ & $2.181\times 10^{-3}$\\
$|y_2|$ & $1.855\times 10^{-3}$ & $1.932\times 10^{-3}$ & $1.073\times 10^{-3}$\\
$|y_3|$ & $3.829\times 10^{-3}$ & $1.009\times 10^{-3}$ & $4.795\times 10^{-3}$\\
$|\mu_S|$ [keV] & $4.286$ & $6.556$ & $7.310$\\
$|\mu_S'|$ [keV] & $1.234$ & $3.320$ & $3.125$\\
\hline
$\sigma(\mu)$ [keV] & $5.3,\,4.7,\,3.2$ & $9.8,\,6.7,\,4.7$ & $10.4,\,6.7,\,6.0$\\
$\sigma(M_D)$ [GeV] & $999,\,623,\,436$ & $924,\,811,\,256$ & $922,\,781,\,325$\\
$\sigma(M_{RS})$ [$10^5$GeV] & $4.0,\,2.7,\,1.3$ & $5.1,\,2.6,\,2.5$ & $8.7,\,6.4,\,2.3$\\
$|M_{RS}|^2/|\mu|$ [GeV] & $3.4\times 10^{15}$ & $6.5\times 10^{15}$ & $5.2\times 10^{15}$\\
\hline
\end{tabular}\end{table}

\begin{table}[H]\centering
\caption{Neutrino Oscillation observables. All statistical pulls are found to satisfy $|\mathrm{pull}|<0.01\sigma$.}\label{tab:obs}
\begin{tabular}{|l|c|c|c|r|}
\hline
Observable & NuFIT NO & A: NO & NuFIT IO & A: IO\\
\hline
$\Delta m^2_{21}$ [$10^{-5}\ \text{eV}^2$] & $7.49$ & $7.490$ & $7.49$ & $7.490$\\
$\Delta m^2_{31}$ [$10^{-3}\ \text{eV}^2$] & $+2.534$ & $+2.534$ & $-2.435$ & $-2.435$\\
$sin^2\theta_{12}$ & $0.307$ & $0.30700$ & $0.308$ & $0.30800$\\
$sin^2\theta_{13}$ & $0.02195$ & $0.02195$ & $0.02224$ & $0.02224$\\
$sin^2\theta_{23}$ & $0.561$ & $0.56100$ & $0.562$ & $0.56199$\\
$\delta_{CP}$ [deg] & $177$ & $177.00$ & $285$ & $285.17$\\
$J_{CP}$ & $1.736\!\times\!10^{-3}$ & $1.737\!\times\!10^{-3}$
        & $-3.226\!\times\!10^{-2}$ & $-3.224\!\times\!10^{-2}$\\
        \hline
\end{tabular}\end{table}

\begin{figure}
        \centering
        \includegraphics[width=0.5\linewidth]{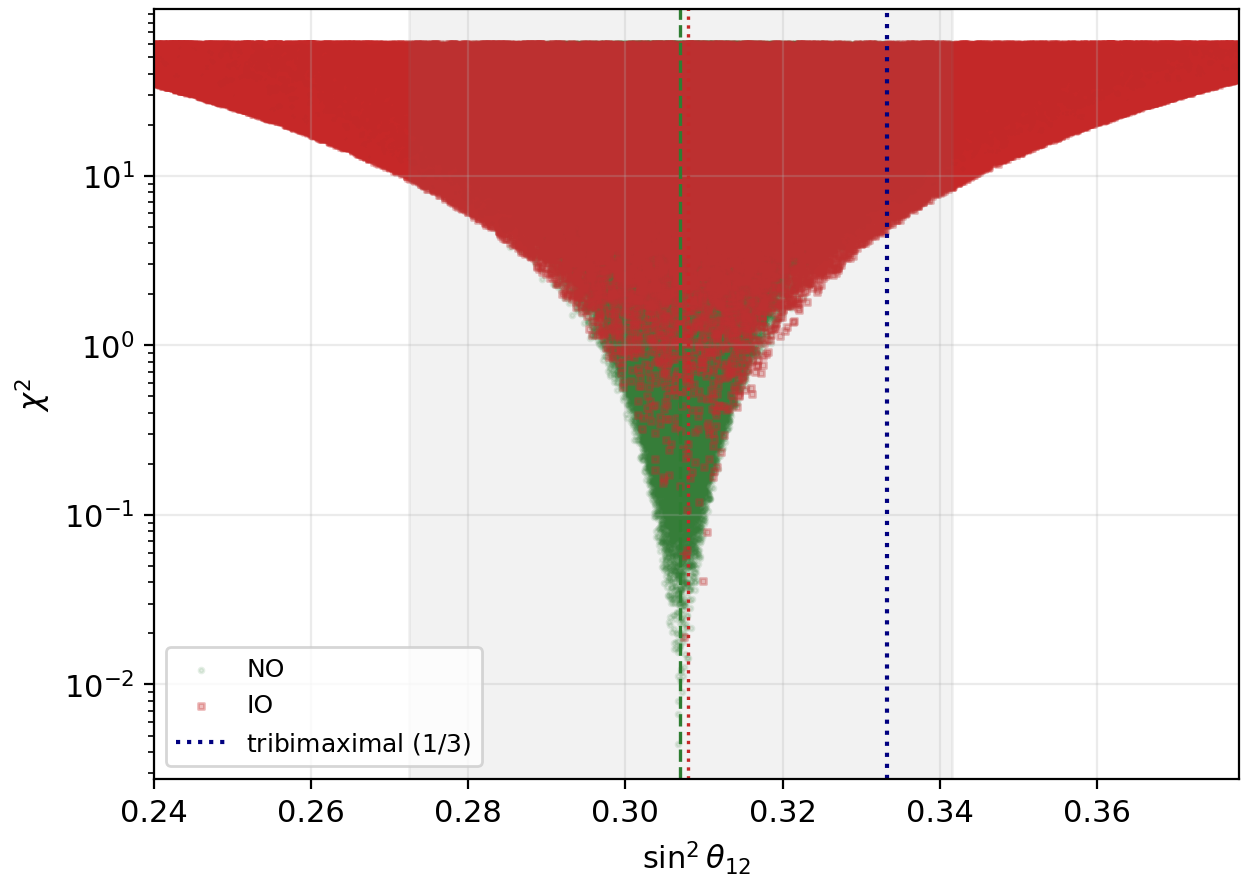}
         \label{fig:placeholder}
        \caption{Plot for $\chi^2$ against $sin^2\theta_{12}$ for normal (green) and inverted (red) ordering. In this plot, the minimum lies at the NuFIT 6.0\cite{Esteban2024NuFIT} value (dashed) and it is clearly distinguished from the tribimaximal value 1/3 (dotted)}.
    \label{fig:one}
\end{figure}

\begin{figure}[H]
    \centering
     \begin{minipage}[t]{0.48\textwidth}
        \centering
        \includegraphics[width=\textwidth]{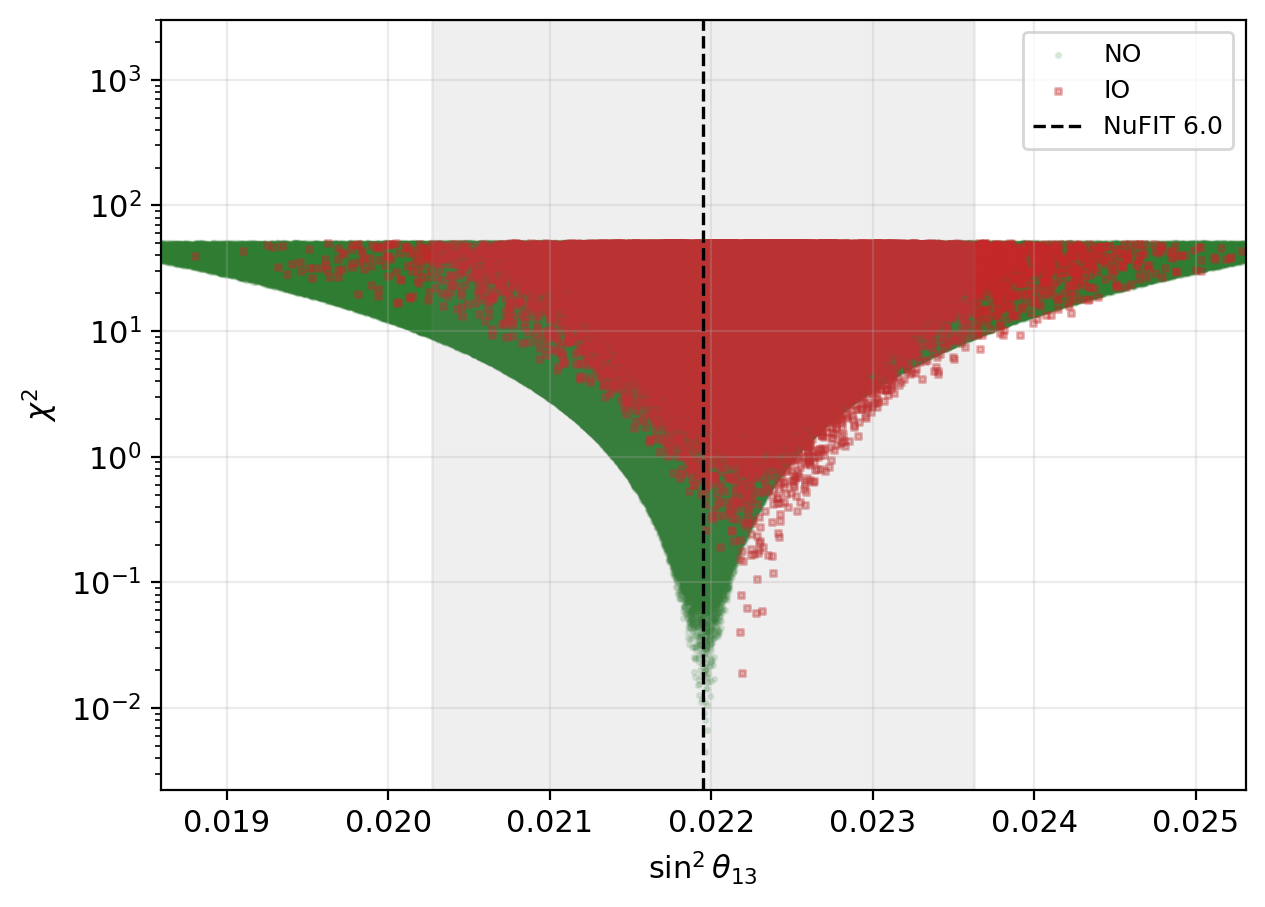}
        \captionof{figure}{The plot for $\chi^2$ against $sin^2\theta_{13}$ for normal (green) and inverted (red) ordering.}
        \label{fig:figure1}
    \end{minipage}
    \hfill
    \begin{minipage}[t]{0.48\textwidth}
        \centering
        \includegraphics[width=\textwidth]{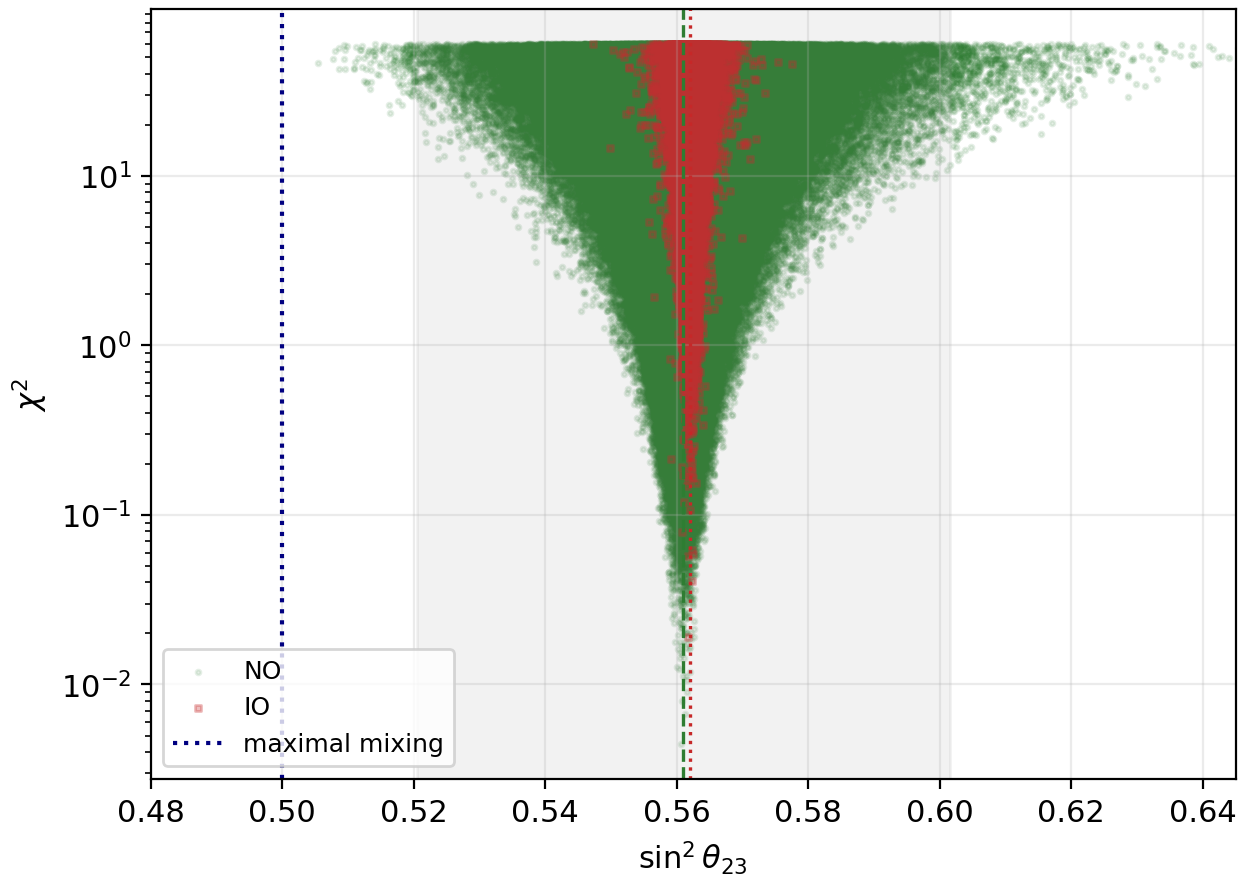}
        \captionof{figure}{The plot $\chi^2$ against $sin^2\theta_{23}$ for normal (green) and inverted (red) ordering. The minimum is displaced from maximal mixing (dotted); the model is adopting the upper octant.}
        \label{fig:figure2}
    \end{minipage}

\end{figure}
    
\begin{table}[H]\centering
\caption{Predicted value of $m_{\beta\beta}$, defined by analytical interval of Eq.~\eqref{eq:env} at the corresponding $\sum m_i$, together with active sterile mixing and non-unitarity of $U_{PMNS}$.}\label{tab:obs}
\begin{tabular}{|l|c|c|c|r|}
\hline
Observable & Loose bound NO & Loose bound IO & Tight bound NO & tight bound IO\\
\hline
$m_1\ \text{meV}$ & $21.68$ & $49.52$ & $2.97$ & ---\\
$m_2\ \text{meV}$ & $23.34$ & $50.27$ & $9.15$ & ---\\
$m_3\ \text{meV}$ & $54.81$ & $4.14$ & $50.43$ & ---\\
$\sum m_i\ \text{meV}$ & $99.8$ & $103.9$ & $62.6$ & $>$ DESI 2024 limit\\
$m_{\beta\beta}\ \text{meV}$ & $6.5$-$22.9$ & $18.3$-$48.7$ & $0.0$-$5.9$
 & KamLAND-Zen $28$-$122$\\
\quad realised value & $(21.0)$ & $(40.9)$ & $(2.93)$ & ---\\
$\sigma_1(\Theta)$ & $5.6\!\times\!10^{-3}$ & $3.6\!\times\!10^{-3}$
 & $3.2\!\times\!10^{-3}$ & $\lesssim\mathcal O(0.1)$\\
$|\eta_{ee}|$ & $4.0\!\times\!10^{-6}$ & $4.0\!\times\!10^{-6}$
 & $1.3\!\times\!10^{-6}$ & $\lesssim10^{-3}$\\

\hline
\end{tabular}\end{table}

\subsection{Neutrino mass spectra}
\begin{figure}[H]
    \centering

    \begin{minipage}[t]{0.48\textwidth}
        \centering
        \includegraphics[width=\textwidth,height=5cm,keepaspectratio]{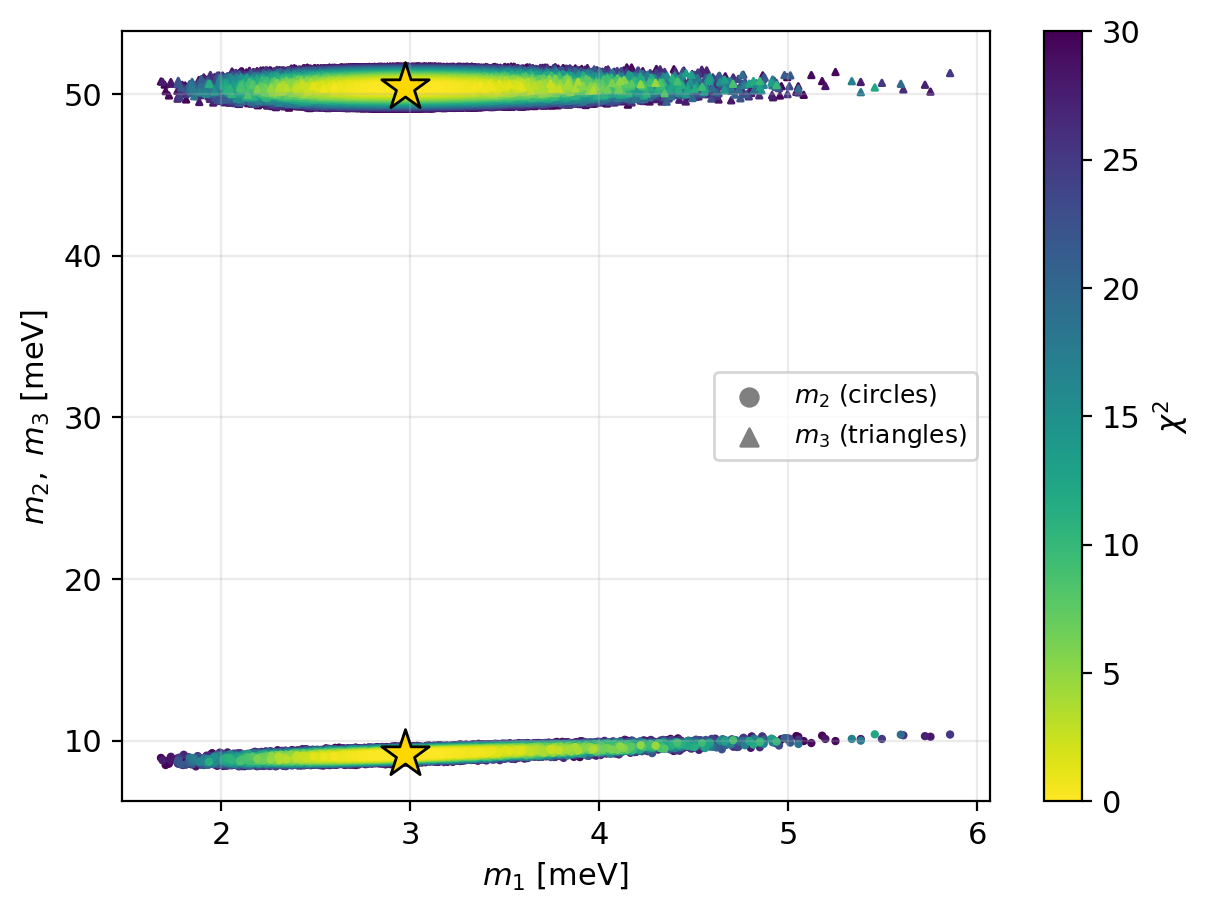}
        \captionof{figure}{The plot for mass eigen states $m_2$ and $m_3$ against $m_1$ for NO, $m_3$ fixed at $50\ \text{meV}$ by $\Delta m_{32}^2$ and is effectively independent of $m_1$ (Pearson $r=0.005)$, while $m_2$ is weakly correlated with $m_1$ $(r=+0,646)$.}
                \label{fig:figure1}
    \end{minipage}
    \hfill
    \begin{minipage}[t]{0.48\textwidth}
        \centering
        \includegraphics[width=\textwidth,height=5cm,keepaspectratio]{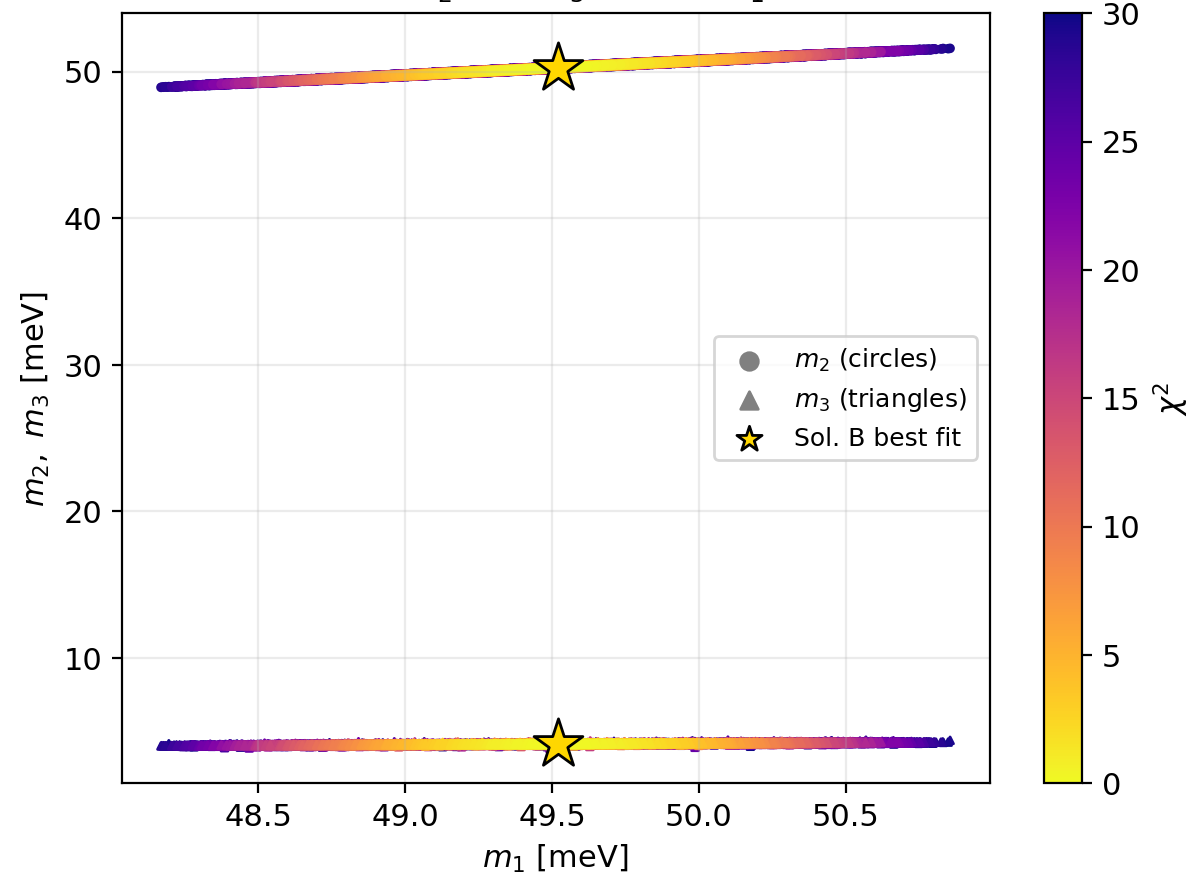}
        \captionof{figure}{The plot for $m_2$ and $m_3$ against $m_1$ for IO, where $m_2$ weakly correlated $m_1$ almost perfectly due to near-degenerate heavy pair ($r=+0.997$) and $m_3$ is the small lightest state independent from $m_1$.}
        \label{fig:figure2}
    \end{minipage}

\end{figure}

\subsection{CP violation: $\delta_{CP}$ as a prediction}
The allowed solution for $\delta_{CP}$ lies within $\delta_{CP}\in [150^{\circ},201^{\circ}]$. There is no solution for $\delta_{CP}$ between $0^{\circ}-90^{\circ}$ and $270^{\circ}-360^{\circ}$. The NuFIT 6.0 \cite{Esteban2024NuFIT} best-fit central value of $177^{\circ}$ lies near the midpoint of the predicted interval. The weak correlations with \(\theta_{12}\) and \(\theta_{13}\) are consistent with the model's flavor structure, in which \(\theta_{23}\) and \(\delta_{CP}\) are associated with the \(\mu\)–\(\tau\) sector.

\begin{figure}[H]
    \centering

    \begin{minipage}[t]{0.48\textwidth}
        \centering
        \includegraphics[width=\textwidth,height=5cm,keepaspectratio]{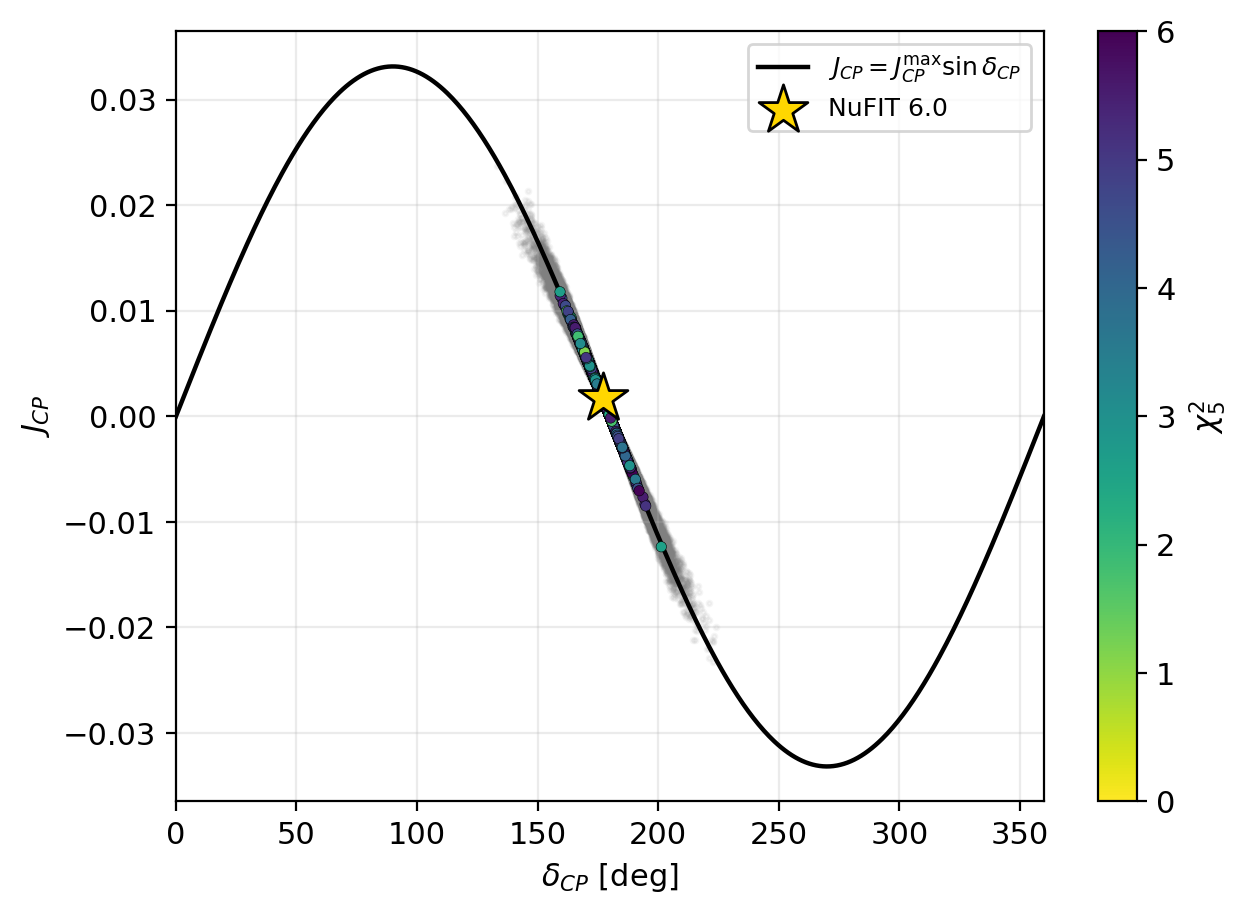}
        \captionof{figure}{The plot for Jarlskog invariant against the Dirac phase for NO. The solid curve is the exact identity
$J_{\mathrm{CP}} = J_{\mathrm{CP}}^{\mathrm{max}}\sin\delta_{\mathrm{CP}}$. The model specifies the clustering near
$\delta_{\mathrm{CP}} \simeq 180^\circ$, giving
$|J_{\mathrm{CP}}| \simeq 1.7 \times 10^{-3}$.}
        \label{fig:figure1}
    \end{minipage}
    \hfill
    \begin{minipage}[t]{0.48\textwidth}
        \centering
        \includegraphics[width=\textwidth,height=5cm,keepaspectratio]{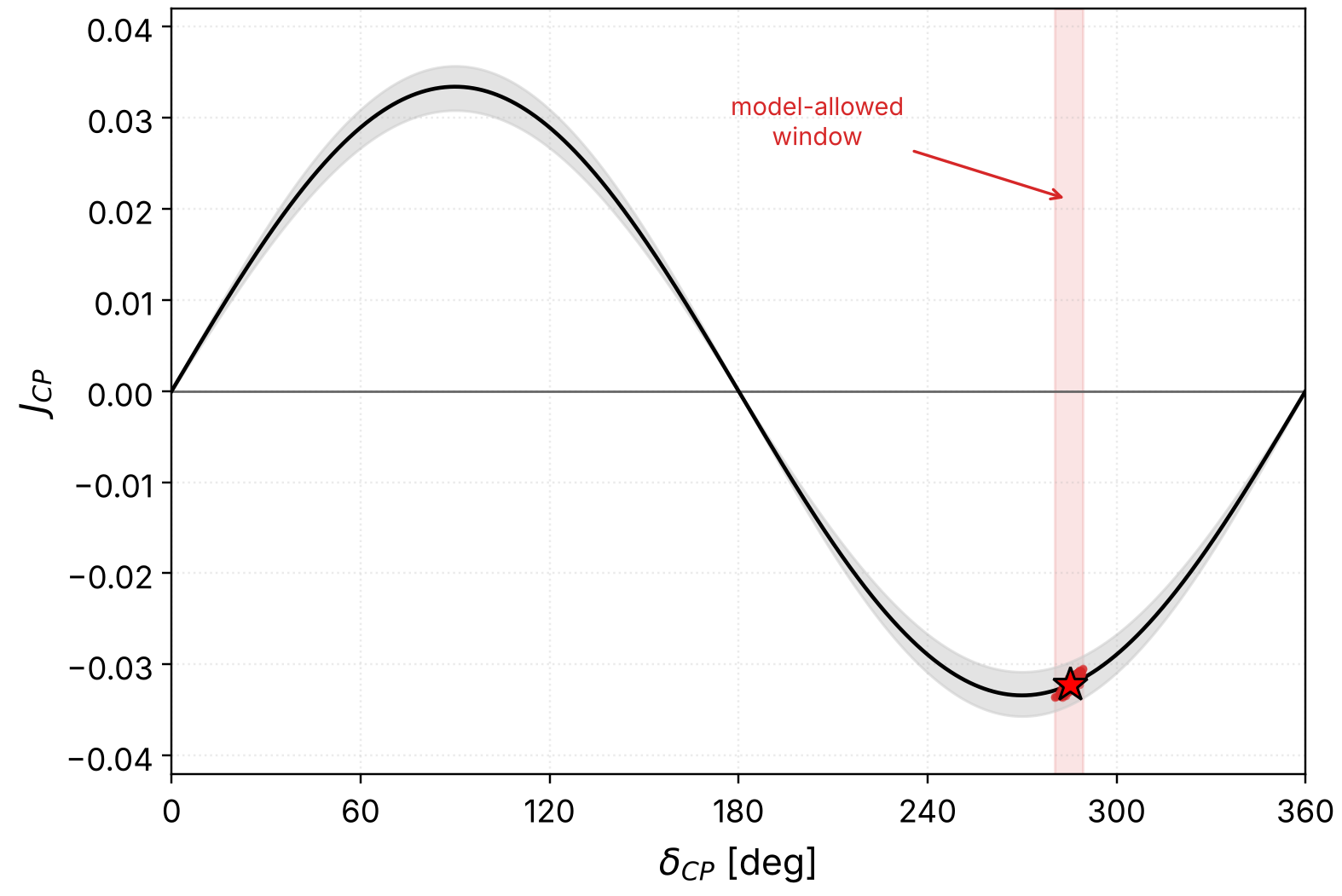}
        \captionof{figure}{The plot for Jarlskog invariant against the Dirac phase for IO. The solid curve is the exact identity
$J_{\mathrm{CP}} = J_{\mathrm{CP}}^{\mathrm{max}}\sin\delta_{\mathrm{CP}}$. The model specifies the clustering near
$\delta_{\mathrm{CP}} \simeq 285^\circ$, allowed region, giving
$|J_{\mathrm{CP}}| \simeq -3.224 \times 10^{-2}$.}

        \label{fig:figure2}
    \end{minipage}

\end{figure}
\subsection{Cosmology dependence of $m_{\beta\beta}$}
\noindent Neutrino oscillation data constrain only two mass-squared differences and the absolute mass scale\cite{Esteban2024NuFIT} $\sum m_i$ is constrained by cosmological data. Choosing $\sum m_i$ and determining the value of the effective Majorana mass, which is given by the two Majorana phases,  results in the allowed range of $m_{\beta\beta}$:
\begin{equation}
m_{\beta\beta}\in\Big[\;0, \big|\,|a|-|b|-|c|\,\big|,\;\;|a|+|b|+|c|\;\Big],
\qquad a=|U_{e1}|^2m_1,\;\; b=|U_{e2}|^2m_2,\;\; c=|U_{e3}|^2m_3 .
\label{eq:env}
\end{equation}
\begin{figure}[H]
    \centering
    \includegraphics[width=0.5\linewidth]{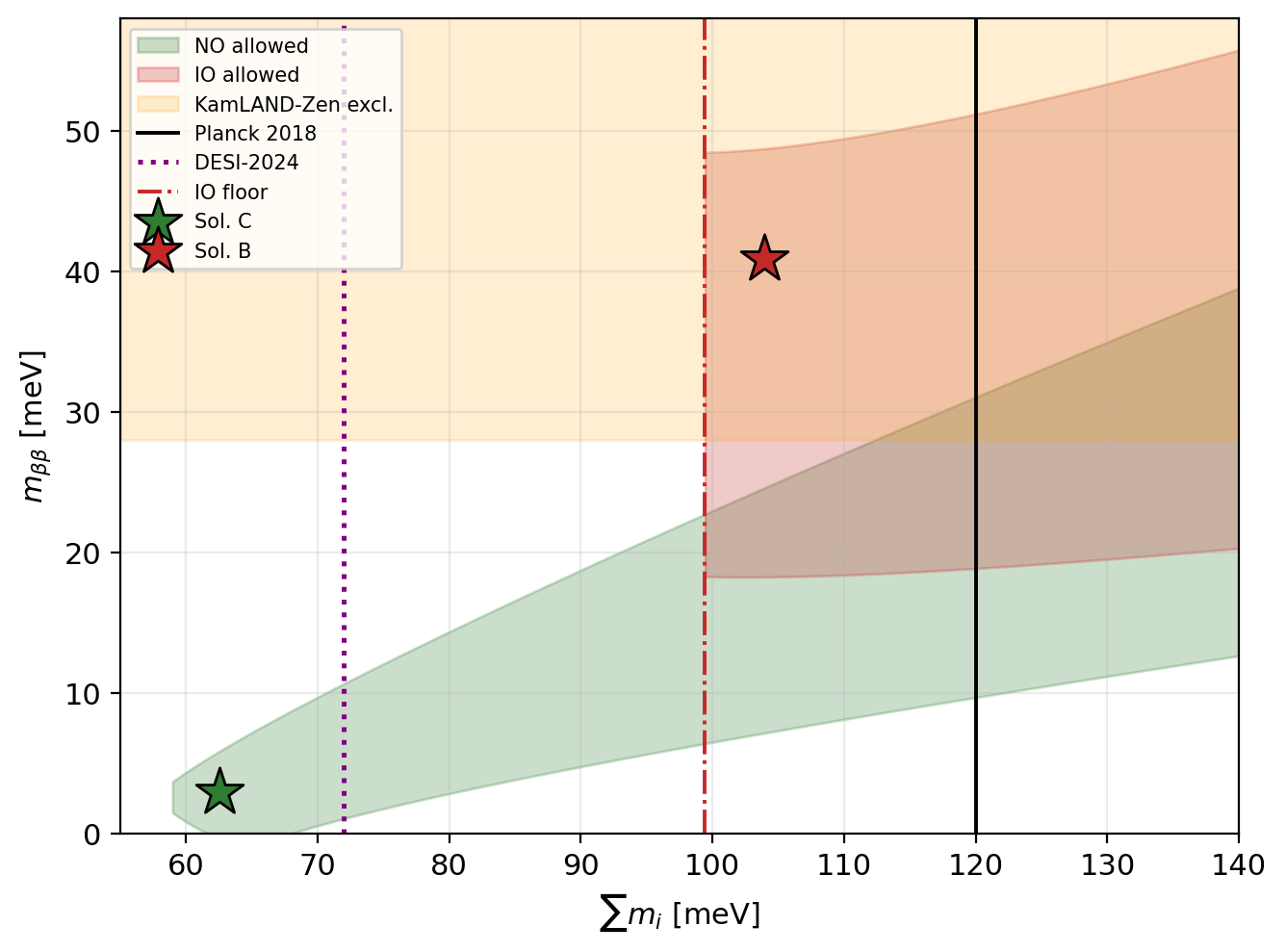}
    \caption{The plot for allowed region for the effective Majorana mass $m_{\beta\beta}$ as a function of $\sum m_i$. Whenever the  $\sum m_i$ is fixed, the three masses are determined, and $m_{\beta\beta}$ spans the interval permitted by the two unconstrained Majorana phases. IO begins only at its kinematic minimum $\sum m_i=99.4$ meV and lies entirely inside the KamLAND-Zen\cite{gando2013limit} excluded band.}
    \label{0011}
    \end{figure}
Here $U_{ei}$ are the elements of the first row of PMNS mixing matrix and the $\sum m_{i}$ are the masses of the light neutrinos. The lower and upper endpoints are simply due to the triangle inequality and correspond to constructive and destructive interference, respectively, between the Majorana phase contributions to the amplitudes. The model is thus shown to reproduce the allowed interval. At $\sum m_i=70.0\ \text{meV}$, two explicit $12$-parameter solutions give $m_{\beta\beta}=0.61$ and $9.55\ \text{meV}$ with oscillation $\chi^2=2.7\times 10^{-2}$ and $1.6\times 10^{-2}$, respectively, both of which satisfy all parameter-range and mass-hierarchy constraints. The analytically predicted interval is $m_{\beta\beta}\in[0.57,9.71],\text{meV}$. A second pair of solutions at $\sum m_i=100,\text{meV}$ exhibits similar behavior. Representative intervals are:
\begin{table}[H]
\centering
\begin{tabular}{| l|c|c|r|}
\hline
$\sum m_i$ [meV] & $62$ & $100$ & $130$\\
\hline
$m_{\beta\beta}^{\rm NO}$ [meV] & $0.0$--$5.6$ & $6.5$--$23.0$ & $11.2$--$35.0$\\
\hline
\end{tabular}\end{table}

\noindent For IO, at the lower bound of $\sum m_i=99.4,\text{meV}$, the predicted effective Majorana mass lies in the interval $m_{\beta\beta}\in[18.3,48.5],\text{meV}$. Consequently, the model does not provide tighter constraints on the effective Majorana mass \(m_{\beta\beta}\) than those obtained in the standard three-neutrino framework. Although the proposed model determines the neutrino masses and mixing angles, the two Majorana phases remain unconstrained by the observables included in the fit. Therefore, \(m_{\beta\beta}\) is not uniquely predicted but can vary within an allowed interval determined by the possible superposition among the three neutrino mass contributions. Quoting a single
$m_{\beta\beta}$ from one best-fit point would be misleading. The physical content is the orange region 
 together with the assumed bound on
$\sum m_i$. For IO, $m_{1,2}\ge\sqrt{|\Delta m^2_{31}|}$, which gives  absolute value
\begin{equation}
\sum m_i\Big|_{\rm IO}^{\min}=99.4\ \text{meV},
\qquad
\sum m_i\Big|_{\rm NO}^{\min}=59.0\ \text{meV}.
\end{equation}
Within the adopted cosmological framework, an upper bound on
$\sum m_i$ below the minimum value permitted by IO
would disfavor IO. In particular, the DESI-2024 data combination has been 
considered\cite{Allali2024DESI}, which yields
$\sum m_i<72\,\mathrm{meV}$, and lies below the minimum mass sum
for IO. However, this conclusion depends on the cosmological
dataset combination and assumptions. Imposing the tighter DESI-2024 value $\sum m_i<72\ \text{meV}$ it is found that $\chi^2_{\rm NO}=1.03\times10^{-9}$ against
$\chi^2_{\rm IO}=2.29\times10^{3}$. From our model we get minimum value $\sum m_i=99.4\ \text{meV}$ for IO against tighter DESI 2024 limit. Thus, our model predicts NO for neutrino mass.
\section{Conclusions}\label{5}
In this paper, a new dynamical ISS(3+1+1+1) neutrino mixing model is constructed within the $U(1)\times A_{4}\times Z_{2}$ flavor symmetry. Where we have derived the invariant Lagrangian for ISS(3+1+1+1) and minimized the scalar $A_4\times Z_2$ potential to determine the neutrino mass ordering dynamically. The charged-lepton mass matrix calculated from the Lagrangian is only diagonal if the flavon field $\phi$ has the vacuum alignment $<\phi>= (v_{\phi},v_{\phi},v_{\phi})$. The vacuum interpolation defined in recovers both the normal-ordering and inverted-ordering limits. The Dirac mass matrix $M_{D}(\epsilon)$ is circulant at $\epsilon=0$, independent form $\epsilon$ and gives two component $(M_{D})_{12}=(M_{D})_{31}=0$ at $\epsilon=1$. The Majorana sterile neutrino mass matrix $\mu$ and heavy Dirac neutrino mass matrix $M_{RS}$ possess a two-parameter $A_4$ structure; $\mu$ and $M_{RS}$ reduce to two-parameter $A_4$ textures, giving twelve real free parameters in total. A democratic vacuum alignment of the flavon field $\zeta$ preserve the residual $S_3$ symmetry and therefore leads to tribimaximal mixing, which has been experimentally excluded. This can be resolved with the use of alternative alignment $<\zeta>= (v_{\zeta},0,0)$ which breaks the residual symmetry and therefore can be consistent with experimental neutrino mixing data. The neutrino mass ordering is governed dynamically by the sign of \(\lambda_3-\lambda_3^{\mathrm{crit}}\) in the scalar potential, making \(\epsilon\) a dynamically generated output rather than an externally specified input. The predicted neutrino observables such as mass-squared differences, mixing angles, sum of neutrino masses $\sum m_i$, Dirac CP-violating phase and Jarlskog invariant are in excellent agreement with the NuFIT 6.0 results under both loose and tight bound constraints, satisfying the statistical \(|\mathrm{pull}|<0.01\sigma\). The model strongly favors normal mass ordering (NO), through minimisation of the dynamical scalar potential of the flavon field \(\phi_v\).\\
\noindent Furthermore, the CP-violating Dirac phase is predicted to be in the range of $\delta_{CP}\in[150^\circ,201^\circ]$ and an atmospheric mixing parameter of \(\sin^2\theta_{23}\simeq 0.55\text{--}0.56\), which corresponds to the upper octant. Future long-baseline experiments, particularly DUNE and Hyper-Kamiokande, may have the potential to test these predictions and these experiments may constrain the corresponding neutrino oscillation parameters. The predicted neutrinoless double-beta decay (\(0\nu\beta\beta\)) effective mass map out an allowed region, rather than a unique point, because the Majorana phases are independent of the neutrino observables. IO is excluded by the tighter DESI-2024 cosmological limit $\sum m_{i}=72\ \text{meV}$, which constrains lower limit of the sum of neutrino masses \(\sum m_i\geq 99.4\,\text{meV}\) for IO. According to our model, the predicted total neutrino mass and the effective Majorana mass are $\sum m_{i}\simeq 62\ \text{meV}$ and  $m_{\beta\beta}\in [0,5.6]\ \text{meV}$, respectively; therefore, the model favors (NO). Interestingly, this predicted region of parameter space lies below the prospective sensitivity of LEGEND-1000\cite{cattadori2021gerda} and close to the ultimate sensitivity limit of nEXO\cite{adhikari2022nexo}. Consequently, Future cosmological constraints on $\sum m_i$ may thus form one of the strongest probes of our model.


\bibliographystyle{unsrtnat}
\bibliography{references}
 
\newpage
\section{Appendix}

\begin{table}[htbp]\centering\small
\caption{Fitting parameters, Solution A: NO, loose cosmological limit ($\sum m_i<120$~meV). The connecting
 factor is fixed by the scalar potential to $\epsilon=0$; $\chi^2_{\min}=2.732\times10^{-8}$.}
\label{tab:paramsA}
\renewcommand{\arraystretch}{1.3}
\begin{tabular}{|l|c|c|c|c|}\hline
\textbf{Parameter} & $\mathbf{Re}$ & $\mathbf{Im}$ & \textbf{Modulus} & \textbf{Phase [deg]} \\ \hline
$ m_{RS}\ [\mathrm{GeV}] $ & \multicolumn{2}{c|}{$1.33265821\times10^{5}$} & $1.33265821\times10^{5}$ & $+0.0000$ \\ \hline
$ \kappa $ & \multicolumn{2}{c|}{$-2.00923284$} & $2.00923284$ & $+180.0000$ \\ \hline
$ y_1 $ & $-9.11080759\times10^{-4}$ & $-4.12228061\times10^{-4}$ & $1.00000006\times10^{-3}$ & $-155.6551$ \\ \hline
$ y_2 $ & $-1.73604910\times10^{-3}$ & $6.53620671\times10^{-4}$ & $1.85501657\times10^{-3}$ & $+159.3687$ \\ \hline
$ y_3 $ & $3.26739197\times10^{-3}$ & $1.99561269\times10^{-3}$ & $3.82861859\times10^{-3}$ & $+31.4152$ \\ \hline
$ \mu_S\ [\mathrm{GeV}] $ & $-2.29812123\times10^{-6}$ & $3.61815211\times10^{-6}$ & $4.28630212\times10^{-6}$ & $+122.4223$ \\ \hline
$ \mu_S'\ [\mathrm{GeV}] $ & $-1.17712927\times10^{-6}$ & $3.71013862\times10^{-7}$ & $1.23421417\times10^{-6}$ & $+162.5059$ \\ \hline
\end{tabular}\end{table}

\begin{table}[htbp]\centering\small
\caption{Fitting parameters, Solution B: IO, loose cosmological limit ($\sum m_i<120$~meV). The connecting
 factor is fixed by the scalar potential to $\epsilon=1$; $\chi^2_{\min}=3.970\times10^{-5}$.}
\label{tab:paramsB}
\renewcommand{\arraystretch}{1.3}
\begin{tabular}{|l|c|c|c|c|}\hline
\textbf{Parameter} & $\mathbf{Re}$ & $\mathbf{Im}$ & \textbf{Modulus} & \textbf{Phase [deg]} \\ \hline
$ m_{RS}\ [\mathrm{GeV}] $ & \multicolumn{2}{c|}{$1.28511379\times10^{5}$} & $1.28511379\times10^{5}$ & $+0.0000$ \\ \hline
$ \kappa $ & \multicolumn{2}{c|}{$-2.96284627$} & $2.96284627$ & $+180.0000$ \\ \hline
$ y_1 $ & $1.02651075\times10^{-3}$ & $7.28347198\times10^{-5}$ & $1.02909146\times10^{-3}$ & $+4.0585$ \\ \hline
$ y_2 $ & $1.82722280\times10^{-3}$ & $6.28150364\times10^{-4}$ & $1.93217909\times10^{-3}$ & $+18.9716$ \\ \hline
$ y_3 $ & $-1.25779266\times10^{-4}$ & $1.00112009\times10^{-3}$ & $1.00899052\times10^{-3}$ & $+97.1610$ \\ \hline
$ \mu_S\ [\mathrm{GeV}] $ & $2.78419496\times10^{-6}$ & $-5.93539842\times10^{-6}$ & $6.55596645\times10^{-6}$ & $-64.8695$ \\ \hline
$ \mu_S'\ [\mathrm{GeV}] $ & $2.24352389\times10^{-6}$ & $-2.44744890\times10^{-6}$ & $3.32015144\times10^{-6}$ & $-47.4892$ \\ \hline
\end{tabular}\end{table}

\begin{table}[htbp]\centering\small
\caption{Fitting parameters, Solution C: NO, tight cosmological limit ($\sum m_i<72$~meV). The connecting
 factor is fixed by the scalar potential to $\epsilon=0$; $\chi^2_{\min}=1.026\times10^{-9}$.}
\label{tab:paramsC}
\renewcommand{\arraystretch}{1.3}
\begin{tabular}{|l|c|c|c|c|}\hline
\textbf{Parameter} & $\mathbf{Re}$ & $\mathbf{Im}$ & \textbf{Modulus} & \textbf{Phase [deg]} \\ \hline
$ m_{RS}\ [\mathrm{GeV}] $ & \multicolumn{2}{c|}{$3.18617067\times10^{5}$} & $3.18617067\times10^{5}$ & $+0.0000$ \\ \hline
$ \kappa $ & \multicolumn{2}{c|}{$-1.73175939$} & $1.73175939$ & $+180.0000$ \\ \hline
$ y_1 $ & $-5.18136377\times10^{-4}$ & $-2.11817848\times10^{-3}$ & $2.18062958\times10^{-3}$ & $-103.7454$ \\ \hline
$ y_2 $ & $-1.03872404\times10^{-3}$ & $-2.68275460\times10^{-4}$ & $1.07280910\times10^{-3}$ & $-165.5185$ \\ \hline
$ y_3 $ & $4.79152597\times10^{-3}$ & $-1.76070917\times10^{-4}$ & $4.79475986\times10^{-3}$ & $-2.1045$ \\ \hline
$ \mu_S\ [\mathrm{GeV}] $ & $-7.27569394\times10^{-6}$ & $7.08277956\times10^{-7}$ & $7.31008755\times10^{-6}$ & $+174.4399$ \\ \hline
$ \mu_S'\ [\mathrm{GeV}] $ & $9.63336523\times10^{-7}$ & $-2.97282404\times10^{-6}$ & $3.12501201\times10^{-6}$ & $-72.0452$ \\ \hline
\end{tabular}\end{table}
\newpage

 Units of mass matrics $M_D$, $M_RS$, $\mu$ in GeV ; $m_\nu$ in eV ; $U_{PMNS}$ dimensionless

\subsection*{Charged lepton mass matrix}
The vaccum alignment $\langle\Phi\rangle=(v_\Phi,v_\Phi,v_\Phi)$ gives charged lepton mass matrix $M_\ell$ diagonal,
\begin{equation}
M_\ell=\frac{v_Hv_\Phi v_\eta}{\Lambda^{2}}\,\mathrm{diag}(y_e,y_\mu,y_\tau)
=\mathrm{diag}\big(5.110\!\times\!10^{-4},\;1.057\!\times\!10^{-1},\;1.777\big)\ \mathrm{GeV}.
\end{equation}

\subsection*{Solution A: NO, loose cosmological limit ($\epsilon=0$)}

\begin{equation}
\resizebox{\textwidth}{!}{$\displaystyle
M_D=\begin{pmatrix}
-3.333\!\times\!10^{1}+5.377\!\times\!10^{2}i & 1.343\!\times\!10^{2}-1.735\!\times\!10^{2}i & 4.310\!\times\!10^{2}-3.930\!\times\!10^{1}i \\[1pt]
1.343\!\times\!10^{2}-1.735\!\times\!10^{2}i & 4.310\!\times\!10^{2}-3.930\!\times\!10^{1}i & -3.333\!\times\!10^{1}+5.377\!\times\!10^{2}i \\[1pt]
4.310\!\times\!10^{2}-3.930\!\times\!10^{1}i & -3.333\!\times\!10^{1}+5.377\!\times\!10^{2}i & 1.343\!\times\!10^{2}-1.735\!\times\!10^{2}i
\end{pmatrix}\ \mathrm{GeV}
$}
\end{equation}

\begin{equation}
M_{RS}=\begin{pmatrix}
2.665\!\times\!10^{5} & 0 & 0 \\[1pt]
0 & 0 & -4.010\!\times\!10^{5} \\[1pt]
0 & 1.345\!\times\!10^{5} & 0
\end{pmatrix}\ \mathrm{GeV}
\end{equation}

\begin{equation}
\resizebox{\textwidth}{!}{$\displaystyle
\mu=\begin{pmatrix}
-2.298\!\times\!10^{-6}+3.618\!\times\!10^{-6}i & -1.177\!\times\!10^{-6}+3.710\!\times\!10^{-7}i & 0 \\[1pt]
-1.177\!\times\!10^{-6}+3.710\!\times\!10^{-7}i & 0 & -2.298\!\times\!10^{-6}+3.618\!\times\!10^{-6}i \\[1pt]
0 & -2.298\!\times\!10^{-6}+3.618\!\times\!10^{-6}i & -1.177\!\times\!10^{-6}+3.710\!\times\!10^{-7}i
\end{pmatrix}\ \mathrm{GeV}
$}
\end{equation}

\begin{equation}
\resizebox{\textwidth}{!}{$\displaystyle
m_\nu=\begin{pmatrix}
8.261\!\times\!10^{-3}-1.931\!\times\!10^{-2}i & 3.053\!\times\!10^{-3}-8.629\!\times\!10^{-3}i & -4.028\!\times\!10^{-3}-2.611\!\times\!10^{-3}i \\[1pt]
3.053\!\times\!10^{-3}-8.629\!\times\!10^{-3}i & 2.581\!\times\!10^{-2}+2.693\!\times\!10^{-2}i & -2.008\!\times\!10^{-2}+3.194\!\times\!10^{-3}i \\[1pt]
-4.028\!\times\!10^{-3}-2.611\!\times\!10^{-3}i & -2.008\!\times\!10^{-2}+3.194\!\times\!10^{-3}i & 3.385\!\times\!10^{-2}-4.872\!\times\!10^{-4}i
\end{pmatrix}\ \mathrm{eV}
$}
\end{equation}

\begin{equation}
\resizebox{\textwidth}{!}{$\displaystyle
U=\begin{pmatrix}
-6.261\!\times\!10^{-1}-5.345\!\times\!10^{-1}i & 5.282\!\times\!10^{-1}+1.460\!\times\!10^{-1}i & -1.022\!\times\!10^{-1}-1.072\!\times\!10^{-1}i \\[1pt]
-2.596\!\times\!10^{-1}+9.058\!\times\!10^{-2}i & -4.290\!\times\!10^{-1}+4.379\!\times\!10^{-1}i & -7.069\!\times\!10^{-1}+2.212\!\times\!10^{-1}i \\[1pt]
-4.960\!\times\!10^{-1}-2.489\!\times\!10^{-2}i & -5.300\!\times\!10^{-1}+2.075\!\times\!10^{-1}i & 6.514\!\times\!10^{-1}+7.125\!\times\!10^{-2}i
\end{pmatrix}
$}
\end{equation}

\begin{equation}
|U|=\begin{pmatrix}
8.233\!\times\!10^{-1} & 5.480\!\times\!10^{-1} & 1.482\!\times\!10^{-1} \\[1pt]
2.749\!\times\!10^{-1} & 6.130\!\times\!10^{-1} & 7.407\!\times\!10^{-1} \\[1pt]
4.966\!\times\!10^{-1} & 5.692\!\times\!10^{-1} & 6.553\!\times\!10^{-1}
\end{pmatrix}
\end{equation}

\noindent Here $m_1=21.6769$, $m_2=23.3407$, $m_3=54.8077$~meV, $\sum m_i=99.825$~meV, $m_{\beta\beta}=21.0040$~meV, $\delta_{CP}=177.00^\circ$ and $J_{CP}=+1.7371\times 10^{-3}$.

\subsection*{Solution B: IO, loose cosmological limit ($\epsilon=1$)}

\begin{equation}
\resizebox{\textwidth}{!}{$\displaystyle
M_D=\begin{pmatrix}
5.745\!\times\!10^{2}+3.675\!\times\!10^{2}i & 0 & 3.311\!\times\!10^{2}-3.453\!\times\!10^{2}i \\[1pt]
1.435\!\times\!10^{2}-6.769\!\times\!10^{1}i & -4.646\!\times\!10^{2}-1.141\!\times\!10^{2}i & -2.048\!\times\!10^{1}+1.630\!\times\!10^{2}i \\[1pt]
0 & -4.951\!\times\!10^{2}+5.760\!\times\!10^{2}i & -1.304\!\times\!10^{2}-9.041\!\times\!10^{1}i
\end{pmatrix}\ \mathrm{GeV}
$}
\end{equation}

\begin{equation}
M_{RS}=\begin{pmatrix}
2.570\!\times\!10^{5} & 0 & 0 \\[1pt]
0 & 0 & -5.093\!\times\!10^{5} \\[1pt]
0 & 2.522\!\times\!10^{5} & 0
\end{pmatrix}\ \mathrm{GeV}
\end{equation}

\begin{equation}
\resizebox{\textwidth}{!}{$\displaystyle
\mu=\begin{pmatrix}
2.784\!\times\!10^{-6}-5.935\!\times\!10^{-6}i & 2.244\!\times\!10^{-6}-2.447\!\times\!10^{-6}i & 0 \\[1pt]
2.244\!\times\!10^{-6}-2.447\!\times\!10^{-6}i & 0 & 2.784\!\times\!10^{-6}-5.935\!\times\!10^{-6}i \\[1pt]
0 & 2.784\!\times\!10^{-6}-5.935\!\times\!10^{-6}i & 2.244\!\times\!10^{-6}-2.447\!\times\!10^{-6}i
\end{pmatrix}\ \mathrm{GeV}
$}
\end{equation}

\begin{equation}
\resizebox{\textwidth}{!}{$\displaystyle
m_\nu=\begin{pmatrix}
3.815\!\times\!10^{-2}+1.477\!\times\!10^{-2}i & 4.838\!\times\!10^{-3}-2.177\!\times\!10^{-2}i & -1.489\!\times\!10^{-2}-5.292\!\times\!10^{-3}i \\[1pt]
4.838\!\times\!10^{-3}-2.177\!\times\!10^{-2}i & 1.452\!\times\!10^{-2}-1.086\!\times\!10^{-3}i & 5.182\!\times\!10^{-3}-1.999\!\times\!10^{-2}i \\[1pt]
-1.489\!\times\!10^{-2}-5.292\!\times\!10^{-3}i & 5.182\!\times\!10^{-3}-1.999\!\times\!10^{-2}i & -2.724\!\times\!10^{-2}-4.686\!\times\!10^{-3}i
\end{pmatrix}\ \mathrm{eV}
$}
\end{equation}

\begin{equation}
\resizebox{\textwidth}{!}{$\displaystyle
U=\begin{pmatrix}
7.690\!\times\!10^{-1}-2.918\!\times\!10^{-1}i & 5.302\!\times\!10^{-1}+1.415\!\times\!10^{-1}i & -7.406\!\times\!10^{-2}+1.294\!\times\!10^{-1}i \\[1pt]
-3.030\!\times\!10^{-1}+2.639\!\times\!10^{-1}i & 5.218\!\times\!10^{-1}+1.298\!\times\!10^{-1}i & 5.875\!\times\!10^{-1}+4.521\!\times\!10^{-1}i \\[1pt]
-1.439\!\times\!10^{-1}-3.758\!\times\!10^{-1}i & 1.487\!\times\!10^{-2}+6.400\!\times\!10^{-1}i & 3.603\!\times\!10^{-1}-5.463\!\times\!10^{-1}i
\end{pmatrix}
$}
\end{equation}

\begin{equation}
|U|=\begin{pmatrix}
8.226\!\times\!10^{-1} & 5.488\!\times\!10^{-1} & 1.491\!\times\!10^{-1} \\[1pt]
4.018\!\times\!10^{-1} & 5.377\!\times\!10^{-1} & 7.413\!\times\!10^{-1} \\[1pt]
4.024\!\times\!10^{-1} & 6.401\!\times\!10^{-1} & 6.544\!\times\!10^{-1}
\end{pmatrix}
\end{equation}

\noindent Here $m_1=49.5197$, $m_2=50.2703$, $m_3=4.1354$~meV, $\sum m_i=103.925$~meV, $m_{\beta\beta}=40.9083$~meV, $\delta_{CP}=285.17^\circ$ and $J_{CP}=-3.2236\times 10^{-2}$.

\subsection*{Solution C: NO, tight cosmological limit ($\epsilon=0$)}

\begin{equation}
\resizebox{\textwidth}{!}{$\displaystyle
M_D=\begin{pmatrix}
4.419\!\times\!10^{2}-1.160\!\times\!10^{2}i & 8.475\!\times\!10^{1}-3.012\!\times\!10^{2}i & 2.535\!\times\!10^{2}+3.885\!\times\!10^{2}i \\[1pt]
8.475\!\times\!10^{1}-3.012\!\times\!10^{2}i & 2.535\!\times\!10^{2}+3.885\!\times\!10^{2}i & 4.419\!\times\!10^{2}-1.160\!\times\!10^{2}i \\[1pt]
2.535\!\times\!10^{2}+3.885\!\times\!10^{2}i & 4.419\!\times\!10^{2}-1.160\!\times\!10^{2}i & 8.475\!\times\!10^{1}-3.012\!\times\!10^{2}i
\end{pmatrix}\ \mathrm{GeV}
$}
\end{equation}

\begin{equation}
M_{RS}=\begin{pmatrix}
6.372\!\times\!10^{5} & 0 & 0 \\[1pt]
0 & 0 & -8.704\!\times\!10^{5} \\[1pt]
0 & 2.332\!\times\!10^{5} & 0
\end{pmatrix}\ \mathrm{GeV}
\end{equation}

\begin{equation}
\resizebox{\textwidth}{!}{$\displaystyle
\mu=\begin{pmatrix}
-7.276\!\times\!10^{-6}+7.083\!\times\!10^{-7}i & 9.633\!\times\!10^{-7}-2.973\!\times\!10^{-6}i & 0 \\[1pt]
9.633\!\times\!10^{-7}-2.973\!\times\!10^{-6}i & 0 & -7.276\!\times\!10^{-6}+7.083\!\times\!10^{-7}i \\[1pt]
0 & -7.276\!\times\!10^{-6}+7.083\!\times\!10^{-7}i & 9.633\!\times\!10^{-7}-2.973\!\times\!10^{-6}i
\end{pmatrix}\ \mathrm{GeV}
$}
\end{equation}

\begin{equation}
\resizebox{\textwidth}{!}{$\displaystyle
m_\nu=\begin{pmatrix}
2.069\!\times\!10^{-4}+2.927\!\times\!10^{-3}i & -2.319\!\times\!10^{-3}-1.577\!\times\!10^{-3}i & -8.021\!\times\!10^{-3}-2.476\!\times\!10^{-3}i \\[1pt]
-2.319\!\times\!10^{-3}-1.577\!\times\!10^{-3}i & 2.460\!\times\!10^{-2}+1.863\!\times\!10^{-2}i & 1.851\!\times\!10^{-2}-1.116\!\times\!10^{-2}i \\[1pt]
-8.021\!\times\!10^{-3}-2.476\!\times\!10^{-3}i & 1.851\!\times\!10^{-2}-1.116\!\times\!10^{-2}i & -2.009\!\times\!10^{-3}-2.407\!\times\!10^{-2}i
\end{pmatrix}\ \mathrm{eV}
$}
\end{equation}

\begin{equation}
\resizebox{\textwidth}{!}{$\displaystyle
U=\begin{pmatrix}
8.183\!\times\!10^{-1}+9.012\!\times\!10^{-2}i & -2.581\!\times\!10^{-1}+4.834\!\times\!10^{-1}i & -9.320\!\times\!10^{-2}+1.152\!\times\!10^{-1}i \\[1pt]
-1.995\!\times\!10^{-1}-1.891\!\times\!10^{-1}i & -5.508\!\times\!10^{-1}+2.690\!\times\!10^{-1}i & 7.051\!\times\!10^{-1}-2.270\!\times\!10^{-1}i \\[1pt]
-1.482\!\times\!10^{-1}+4.740\!\times\!10^{-1}i & 4.363\!\times\!10^{-1}+3.655\!\times\!10^{-1}i & 4.423\!\times\!10^{-1}+4.835\!\times\!10^{-1}i
\end{pmatrix}
$}
\end{equation}

\begin{equation}
|U|=\begin{pmatrix}
8.233\!\times\!10^{-1} & 5.480\!\times\!10^{-1} & 1.482\!\times\!10^{-1} \\[1pt]
2.749\!\times\!10^{-1} & 6.130\!\times\!10^{-1} & 7.407\!\times\!10^{-1} \\[1pt]
4.966\!\times\!10^{-1} & 5.692\!\times\!10^{-1} & 6.553\!\times\!10^{-1}
\end{pmatrix}
\end{equation}

\noindent Here $m_1=2.9725$, $m_2=9.1507$, $m_3=50.4265$~meV, $\sum m_i=62.550$~meV, $m_{\beta\beta}=2.9345$~meV, $\delta_{CP}=177.00^\circ$ and $J_{CP}=+1.7358\times 10^{-3}$.

\end{document}
```